\documentclass[a4paper]{article}

\usepackage{amsfonts}
\usepackage{amssymb,amsmath,amsthm,mathrsfs}
\usepackage{geometry}
\usepackage{enumerate}
\usepackage{bbm}
\usepackage{graphicx}
\usepackage{bbm}
\usepackage{xcolor}

\usepackage[colorlinks,
            linkcolor=blue,
            anchorcolor=green,
            citecolor=blue
            ]{hyperref}

\numberwithin{equation}{section}

\newcommand{\D}{\mathbb{U}}
\newcommand{\QD}{\mathrm{QD}}

\title{Backbone exponent and annulus crossing probability for planar percolation}
\author{Pierre Nolin\thanks{City University of Hong Kong}\qquad Wei Qian\thanks{City University of Hong Kong (on leave from CNRS,  Laboratoire de Math\'ematiques d'Orsay)} \qquad Xin Sun\thanks{Beijing International Center for Mathematical Research, Peking University} \qquad Zijie Zhuang
\thanks{Wharton Statistics and Data Science Department, University of Pennsylvania}}

\begin{document}

\maketitle
 
\begin{abstract}
We report the recent derivation of the backbone exponent for 2D percolation. In contrast to previously known exactly solved percolation exponents, the backbone exponent is a transcendental number, which is a root of an elementary equation. We also report an exact formula for the probability that there are two disjoint paths of the same color crossing an annulus. The backbone exponent captures the leading asymptotic, while the other roots of the elementary equation capture the asymptotic of the remaining terms. This suggests that the backbone exponent is part of a conformal field theory (CFT) whose bulk spectrum contains this set of  roots. Our approach is based on the coupling between SLE curves and Liouville quantum gravity (LQG), and the integrability of Liouville CFT that governs the LQG surfaces.  
\end{abstract}
\maketitle 

\section{Introduction}
As a fundamental model of critical phenomena, percolation~\cite{BH57} has broad applications in physics and other natural sciences~\cite{saberi2015recent}. The fractal geometry of critical percolation clusters is a classical topic in statistical physics.
For two-dimensional Bernoulli percolation,   
the exact values of many fractal dimensions are explicitly known. These values were often first discovered in physics and later proved using probabilistic methods.
For instance, the dimensions of the percolation cluster, cluster boundary, and pivotal points (also known as ``red bonds'') are $\frac{91}{48}, \frac{7}{4}$, and $\frac{3}{4}$, respectively. Despite these successful examples, the fractal dimension of the backbone of a percolation cluster remained elusive for a long time. The backbone is the part of a percolation cluster that remains after removing the dangling ends, where electrical current between distant vertices would flow. In this note, we report recent progress on the exact derivation of the backbone exponent and the related annulus crossing probability.

The backbone dimension $D_B$ can be characterized as follows. 
Consider critical Bernoulli percolation on a triangular lattice of small mesh size, where each site is colored black or white with equal probability. Let $A(r,R)$ be the annulus of inner radius $r$ and outer radius $R$. As the mesh size tends to $0$, the probability that there are two disjoint black paths crossing an annulus $A(r,R)$ converges to a limiting probability $p_B(r,R)$. There exists an exponent $x_B$ such that $p_B(r,R)$ decays as $(\frac{r}{R})^{x_B + o(1)}$ as $r/R \to 0$ \cite{BN2011}. Then we have $D_B = 2 - x_B$.
The exponent $x_B$ is known as the backbone exponent, or the monochromatic two-arm exponent.

In general, arm exponents of percolation describe the asymptotic behavior of the annulus crossing probability. The monochromatic $k$-arm exponent corresponds to $k$ disjoint black paths.  For $k\ge 2$, the polychromatic $k$-arm exponents correspond to $k$ disjoint paths, not all of which are of the same color. The one-arm exponent and the polychromatic $k$-arm exponents are explicitly known to be $\frac{5}{48}$ and $\frac{k^2-1}{12}$, respectively~\cite{denNijs1979, SaleurDuplantier-1987, ADA99, SW01, LSW02}. These in particular give the dimensions $\frac{91}{48}, \frac{7}{4}$, and $\frac{3}{4}$ mentioned above.

In our recent work~\cite{NQSZ-backbone}, the exact value of $x_B$ is shown to be the unique solution in $(\frac{1}{4},\frac{2}{3})$ to \begin{equation}\label{eq:backbone-exponent}
    \frac{\sqrt{36 x +3}}{4} + \sin \Big(\frac{2 \pi \sqrt{12 x +1}}{3} \Big) =0.   
\end{equation}
The numerical value is 0.3566668\ldots, which matches well with numerical simulations in~\cite{G99, DBN04, FKZD22}. Using this expression, we show that $x_B$ is a transcendental number, which is surprising since all other known exponents in Bernoulli percolation are rational.

Our derivation is based on the convergence of 2D percolation towards Schramm-Loewner evolution (SLE)~\cite{Sc00} with parameter 6. We focus on site percolation on the triangular lattice because the convergence to SLE$_6$ is established only for this lattice~\cite{Sm01}. This convergence is believed to hold for a broad class of 2D Bernoulli percolation, such as bond percolation on the square lattice.  Another key ingredient in our derivation is the coupling between SLE and Liouville quantum gravity (LQG), which describes the scaling limit of statistical physics models on random triangulations. The quantum gravity method for deriving fractal dimension has for example been applied to the dimension $4/3$ of the Brownian frontier~\cite{MR1666816} which was later proved using SLE in~\cite{MR1879851, LSW-III}. This idea, known as the KPZ relation~\cite{KPZrelation}, was put into a mathematical framework in~\cite{DS-KPZ-invent}.

Compared to~\cite{MR1666816}, the main novelty of our approach is to incorporate the integrability of Liouville conformal field theory, the field theory that governs the random surfaces in LQG~\cite{polyakov1981quantum}, into the study of SLE. This approach was developed in~\cite{AHS21} and has been successfully used in several problems.
In fact, we use this method to derive in \cite{SXZ24-annulus} that $p_B(r,R)$ exactly equals 
\begin{equation} \label{eq:backbone-annulus}
\frac{q^{-\frac{1}{12}}}{\prod_{n=1}^\infty (1- q^{2n})} \sum_{s \in \mathcal{S}} \frac{- \sqrt{3} \sin(\frac{2\pi}{3}  \sqrt{3 s}) \sin(\pi \sqrt{3 s}) }{\cos(\frac{4\pi }{3}  \sqrt{3 s}) + \frac{3 \sqrt{3}}{8 \pi} } q^{s},
\end{equation}
where $\mathcal{S} = \{s\in \mathbb C : \sin(4 \pi \sqrt{\frac{s}{3}}) + \frac{3}{2}\sqrt{s} = 0 \} \setminus \{0, \frac13 \}$ and $q = r/R$. The numerical values for $\mathcal{S}$ are $\{ 0.440,\, 2.194 \pm 0.601i,\, 5.522 \pm 1.269 i, \ldots\}$. This equation is related to~\eqref{eq:backbone-exponent} by $s = x + \frac{1}{12}$. Hence~\eqref{eq:backbone-annulus} provides a physical interpretation of roots in~\eqref{eq:backbone-exponent}.

The analogous annulus crossing probabilities for the one-arm exponent and the alternating $2k$-arm exponents were derived by Cardy~\cite{Car02, Car06}, and can be expressed in a similar form as~\eqref{eq:backbone-annulus}, except that all the exponents and coefficients in the expansion are rational. Our derivation of~\eqref{eq:backbone-annulus} is based on a method developed in~\cite{ARS2022moduli}. The same method also recovers Cardy's result for the one-arm case and the alternating two-arm case.  In this note we present the key ideas in deriving $x_B$ and $p_B(r,R)$. 

\section{Conformal radii of SLE and percolation exponents} 

\begin{figure}[h]
\vspace{-0.2cm}
\begin{center}
\includegraphics[angle=0,width=.93290\linewidth]{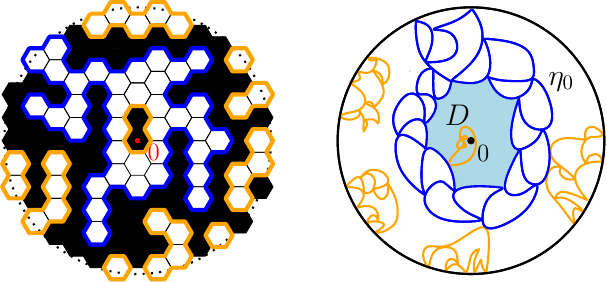}
\end{center}
\vspace{-0.5cm}\caption{\textbf{Left:} Bernoulli percolation on a triangular lattice on the unit disk. The blue loop is the outermost percolation interface surrounding the origin, whose scaling limit is $\eta_0$ on the right. \textbf{Right:} CLE$_6$ on the unit disk. The loops are nested, non-simple, and may touch each other and the boundary. }\label{fig:1}
\vspace{-0.3cm}
\end{figure}

The first step in our derivation of~\eqref{eq:backbone-exponent} is to encode the backbone exponent using the conformal radius of a domain bounded by an SLE$_6$ curve. This type of encoding can be done for various percolation exponents. We first demonstrate this using the example of the one-arm exponent $x_1$. On the triangular lattice with small mesh size, consider Bernoulli site percolation in an approximation of the unit disk, as shown in Figure~\ref{fig:1}, where each site is represented by a hexagon. We set the boundary to be all black, namely, the hexagons not drawn in Figure~\ref{fig:1} are black. Then as the mesh size tends to zero, the interfaces separating the black and white clusters converge to a  collection of loops, which together form the so-called conformal loop ensemble with parameter $\kappa=6$ (CLE$_6$) on the disk~\cite{CN06-full, SheffieldCLE}. Each loop in CLE$_6$ locally looks like the SLE$_6$ curve introduced by Schramm~\cite{Sc00}, which is not self-crossing but has self intersections.  

Let $\eta_0$ be the outermost loop in CLE$_6$ that surrounds the origin. Let $D$ be the complementary connected component of $\eta_0$ containing the origin, as shown in Figure~\ref{fig:1} (right). Let $f$ be a conformal map from the unit disk to $D$ such that $f(0)=0$. The conformal radius ${\rm CR}(0,D)$ of $D$ viewed from the origin $0\in \mathbb C $ is defined as $|f'(0)|$. Let $p(r,1)$ be the probability that there exists a black crossing of $A(r,1)$, which is equivalent to the distance from 0 to the outermost percolation interface being less than $r$. By the Koebe $1/4$ theorem, the ratio between ${\rm CR}(0,D)$ and  the distance from 0 to $\eta_0$ is between $1$ and $4$. 
The one-arm exponent $x_1$ is therefore given by $p(r,1)\approx \mathbb P[{\rm CR}(0,D) \leq r]=r^{x_1+o(1)}$ as $r\to 0$. 
Hence,\begin{equation}\label{eq:one-arm-CR}
x_1=\inf \{x \in \mathbb{R}: \langle {\rm CR}(0,D)^{-x} \rangle =\infty\}.
\end{equation}
See Appendix~\ref{subsec:percolation} for more details. The value $x_1=5/48$
was originally established in \cite{LSW02} by estimating $\mathbb P[{\rm CR}(0,D) \leq r]$. The exact formula $\langle {\rm CR}(0,D)^{-x} \rangle = \frac{1}{2 \cos(\frac{\pi}{3} \sqrt{12x + 1})}$ was further obtained in~\cite{SSW09}, from which one can also extract  $x_1$.

\begin{figure}[h]
\vspace{-0.2cm}
\begin{center}
\includegraphics[angle=0,width=.93290\linewidth]{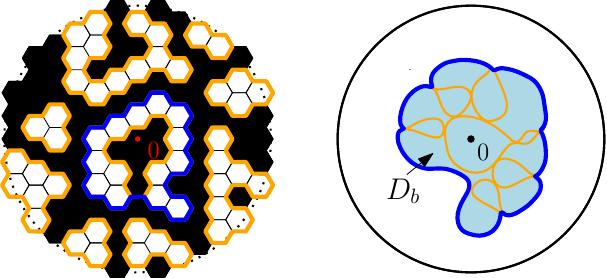}
\end{center}
\vspace{-0.5cm}\caption{\textbf{Left:} Percolation interfaces are colored orange. The blue loop is an example of a filled interface, which encloses all black points that can be surrounded by a white cluster together with an additional black point.
It is the outermost one surrounding the origin. \textbf{Right:} Filled percolation interfaces converge to the outer boundaries of CLE$_6$ loops. The blue loop is the outermost one surrounding 0, and $D_b$ is the domain enclosed by this blue loop.}\label{fig:2}
\vspace{-0.3cm}
\end{figure}

The backbone exponent $x_B$ can be encoded using a similar idea. For each loop in a CLE$_6$ on the unit disk, 
its outer boundary is defined to be the boundary of the unbounded component after removing the loop from the plane. 
Then the outer boundaries of all the CLE$_6$ loops form a random collection of simple loops, each of which looks like an SLE$_\kappa$ curve with $\kappa=\frac83$ \cite{LSW_CR_chordal}. Let $D_b$ be the domain bounded by the outermost loop surrounding the origin in this loop ensemble. Then 
\begin{equation}\label{eq:backbone-CR}
x_B= \inf \{ x \in \mathbb{R}: \langle {\rm CR}(0,D_b)^{-x} \rangle = \infty \}.
\end{equation}
To see~\eqref{eq:backbone-CR}, consider the \textit{filled percolation interfaces}, which are obtained by filling the fjords (i.e., passages with width 1); see the blue loop in Figure~\ref{fig:2} (left). If the distance from 0 to the outermost filled percolation interface is less than $r$, then we need to flip at least two black points to disconnect all black crossings in $A(r, 1)$. By Menger's theorem, there exist two disjoint black crossings in $A(r, 1)$. In fact, the probabilities of these two events share the same exponent as $r\to 0$; see Appendix~\ref{subsec:percolation}. Moreover, these filled percolation interfaces converge to outer boundaries of the CLE$_6$ loops. This gives~\eqref{eq:backbone-CR} similar to~\eqref{eq:one-arm-CR}.

Now the value of $x_B$ specified by~\eqref{eq:backbone-CR} follows from 
the exact formula of $\langle {\rm CR}(0,D_b)^{-x} \rangle$, which is 
\begin{equation}\label{eq:backbone-CR-formula}
\frac{3\sqrt{3}}{4} \sin(\frac{\pi}{2} \sqrt{12x+1}) \Big(\frac{\sqrt{36 x +3}}{4} + \sin (\frac{2 \pi \sqrt{12 x +1}}{3} ) \Big)^{-1}.
\end{equation}
Below we explain how $\langle {\rm CR}(0,D_b)^{-x} \rangle$ and $p_B(r,R)$ arise in the 2D quantum gravity framework. 

\section{The quantum gravity approach}

Consider Bernoulli site percolation on a random triangulation.
We can view the random triangulation as a discrete model for 2D quantum gravity, see e.g.~\cite{AMBJORN1995129,BOULATOV1987379,MR2013797}, and percolation as a conformal matter with central charge $c_M=0$. Assume the random triangulation has the disk topology. 
Then in the continuum limit, the 2D quantum gravity can be described by the Liouville field theory on the disk with central charge $c_L=26-c_M=26$. The action of the theory is 
\begin{equation}\label{eq:liouville-action}
\begin{aligned}
S_L[\phi] = \int_{\D} (\frac{1}{4 \pi} |\nabla \phi|^2 + \mu e^{\gamma \phi}) d^2 x + \int_{\partial \D} (\frac{Q \phi }{2\pi} + \nu e^{\frac{\gamma}{2} \phi}) d l.
\end{aligned}
\end{equation}
The background charge $Q>2$ is related to $c_L$ by $c_L=1+6Q^2$, hence $Q=\sqrt{25/6}$. 
We take $\D$ to be the flat unit disk $\{|z|\leq 1\}$ so that the charge is supported on the boundary. The coupling constant $\gamma\in (0,2)$  is related to $Q$ by $Q=\frac{\gamma}{2}+\frac{2}{\gamma}$, hence $\gamma=\sqrt{8/3}$. 
The bulk cosmological constant $\mu$ and the boundary cosmological constant $\nu$ are allowed to vary. 
Consider the field $\phi$ whose distribution is given by $e^{\frac{\gamma}{2} \phi(1)} \cdot e^{\gamma \phi(0)} \cdot e^{- S_L[\phi]} D\phi$. 
The continuum limit of percolation on a random triangulation of the disk can be described by CLE$_6$ on $\D$ with a random geometric background: the area measure is $e^{\gamma \phi}d^2 x $ on $\D$, and the boundary length measure is $e^{\frac{\gamma}{2} \phi} d l$ on $\partial \D$. Here $0\in \D$ and $1\in \partial \D$ correspond to one bulk point and one boundary point on the triangulation, which are marked to specify how the random surface is conformally parameterized by $\D$. A version of this is rigorously proved in~\cite{HS19}.

From now on we set $\mu=0$ in $S_L[\phi]$ and write it as $S_L^\nu[\phi]$, only allowing $\nu$ to vary. 
Then
\begin{equation} \label{eq:QD(1,0)}
\int_{\phi: \mathbb U\to \mathbb R} F(\int_{\partial \mathbb U} e^{\frac{\gamma}{2}\phi}d\ell)   e^{\gamma \phi(0)} e^{-S_L^\nu [\phi]} D\phi
\end{equation}
is proportional to $\int_0^\infty F(L) e^{-\nu L}  L^{-\frac32} d L $ for any test function $F$~\cite{LQG-disk}. On the discrete side, 
consider a polygon of length $p$.  For each triangulation of it with an interior marked point and $n$ faces we assign weight 
$a^nb^p$. Then for critical $a$ and $b$, as $p\to \infty$ the total weight grows as $p^{-\frac{3}{2}}$.
From both perspectives, we can say that $L^{-\frac{3}{2}}$ is the  partition function  for the random disk in the pure 2D quantum gravity which has boundary length $L$, one interior marked point, and no area constraint~\cite{BM17-brownian-disk}; see Appendix~\ref{subsec:liouville} for further background. We denote such a random disk by $\mathrm{QD}_{1}(L)$. 
We do not mark any boundary point on $\mathrm{QD}_{1}(L)$, hence the boundary insertion $e^{\frac{\gamma}2 \phi(1)}$ does not appear in~\eqref{eq:QD(1,0)}.

\begin{figure}[h]
\vspace{-0.2cm}
\begin{center}
\includegraphics[angle=0,width=.93290\linewidth]{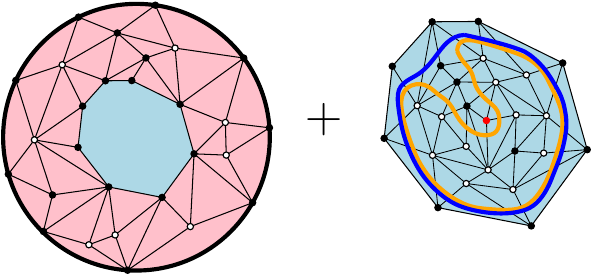}
\end{center}
\vspace{-0.5cm}\caption{Bernoulli site percolation on a random triangulation of the disk. The outermost filled percolation interface colored in blue divides the random triangulation into two parts. These two parts are independent conditioned on the number of vertices along the filled interface. In the continuum limit, this corresponds to~\eqref{eq:key0}.}\label{fig:tri}
\vspace{-0.3cm}
\end{figure}

Suppose $\mathrm{QD}_{1}(L)$ is conformally realized on $\mathbb U$ with the marked point at $0$.
Consider CLE$_6$ on $\mathbb U$ as in Figure~\ref{fig:2} with $D_b$ defined above~\eqref{eq:backbone-CR}, and let $\eta$ be the boundary of $D_b$. 
The curve $\eta$ 
cuts  $\mathrm{QD}_{1}(L)$ into two pieces of random surfaces which are independent 
once conditioned on the length of $\eta$. This independence in the triangulation setting is clear; see Figure~\ref{fig:tri}.
In the continuum, results of this type were pioneered by~\cite{She16a,MR4340069}. A highly nontrivial conclusion from this approach is that  
the surface inside $\eta$ is another copy of $\QD_1$. In fact we can write the following equality of partition functions:
\begin{equation}\label{eq:key0}
    \int_0^\infty Z(L,\ell) \times \ell\times  \ell^{-\frac{3}{2}} d\ell = L^{-\frac{3}{2}},
\end{equation}
where $Z(L,\ell)$ is the partition function of the random surface bounded by $\partial \D$ and $\eta$ with given boundary lengths, and $\ell^{-\frac{3}{2}}$ is the partition function of $\QD_1(\ell)$ which describes the random surface inside $\eta$. The additional factor $\ell$ is the length of $\eta$, that counts the number of ways in which the two surfaces can be glued together, since we do not mark a point on $\eta$. See Appendix~\ref{subsec:liouville} for more details.

We are now  ready to explain the key equation 
\begin{equation}\label{eq:key}
    \int_0^\infty Z(L,\ell) \times \ell\times  \ell^{-\frac32 + a } d\ell = L^{-\frac32 + a} \langle {\rm CR}(0,D_b)^{x(a)} \rangle,
\end{equation}
where $x(a) = -\frac{1}{3}a(a-1)$.
This is a modification of the surface gluing equation~\eqref{eq:key0}. The surface bounded by $\partial \D$ and $\eta$ stays the same, while the surface corresponding to
$\QD_1(\ell)$ is modified to a new random surface with partition function $\ell^{-\frac32 +a}$. To describe this new surface, we note that 
if $e^{\gamma \phi (0)}$ in~\eqref{eq:QD(1,0)} is replaced by $e^{\alpha \phi (0)}$, then \eqref{eq:QD(1,0)} becomes proportional to $ \int_0^\infty F(L) e^{-\nu L}  L^{\frac{2}{\gamma}(\alpha - \gamma) -\frac{3}{2} } d L $. This defines a random surface with partition function $L^{\frac{2}{\gamma}(\alpha - \gamma) -\frac{3}{2}}$ just as how $\QD_1(L)$ is defined from~\eqref{eq:QD(1,0)}.    Taking $\alpha = \gamma + \frac{\gamma}{2} a$ and $\gamma=\sqrt{8/3}$, this defines the random surface in~\eqref{eq:key} with partition function $\ell^{-\frac32 + a }$.

Once the surface corresponding to $\QD_1(\ell)$ in~\eqref{eq:key0} is modified, the surface $\QD_1(L)$ on the right side of~\eqref{eq:key0} 
changes to a surface with partition function  $L^{-\frac32 + a} $ via the same mechanism. The additional factor $\langle {\rm CR}(0,D_b)^{x(a)} \rangle$ in~\eqref{eq:key}
arises because when modifying the bulk insertion $e^{\gamma \phi (0)}$ to $e^{\alpha \phi (0)}$, we assumed that both the smaller surface $\QD_1(\ell)$ and the larger surface $\QD_1(L)$ 
are conformally realized on $\D$. Therefore, we first need to apply a conformal map  $f$ from $D_b$ to $\mathbb U$ while fixing $0$. This results in a factor of $|f'(0)|^{x(a)}$ with 
$x(a) = 2 \Delta_\alpha - 2 \Delta_\gamma$, where $\Delta_\alpha = \frac{\alpha}{2}(Q - \frac{\alpha}{2})$ is the scaling dimension of the bulk insertion $e^{\alpha\phi(0)}$ in Liouville theory. By definition, $|f'(0)|^{-1}={\rm CR}(0,D_b)$, which gives the right side of~\eqref{eq:key}; see Appendix~\ref{subsec:liouville}.

Equations like~\eqref{eq:key} relate the conformal radii of domains bounded by SLE curves with the partition function
of random surfaces in 2D quantum gravity with given lengths. As demonstrated in~\cite{AHS21}, in certain cases, the partition function of the random surfaces can be exactly solved using the structure constants of boundary Liouville theory~\cite{RZ22}, hence such relations provide the exact formula for the conformal radii. However, since the surface corresponding to $Z(L,\ell)$ has a random conformal modulus, the method in~\cite{AHS21}  does not readily give a formula for $Z(L,\ell)$. A key step in~\cite{NQSZ-backbone} to derive~\eqref{eq:backbone-exponent}
is to find an effective variant of  ${\rm CR}(0,D_b)$ such that it still encodes  the backbone exponent; and moreover, the analog of $Z(L,\ell)$ is solvable using the method in~\cite{AHS21}. The strategy is to consider a chain of adjacent loops connecting the disk boundary to the boundary of $D_b$, which forms a sequence of shrinking domains whose conformal radii can be solved step by step. The precise implementation of this  strategy involves a Poisson point process of a special variant of SLE$_6$ called the SLE$_{6}$ bubble \cite{MSW2017}. The detail can be found in~\cite[Section~2]{NQSZ-backbone}.

The ratio between ${\rm CR}(0,D_b)$ and its effective variant in~\cite{NQSZ-backbone} is another conformal radius that can be solved using the method from~\cite{AHS21}. This gives~\eqref{eq:backbone-CR-formula}, as done in~\cite{SXZ24-annulus}. Now $Z(L,\ell)$ can be solved from~\eqref{eq:backbone-CR-formula} and~\eqref{eq:key}. This is the starting point of the derivation of~\eqref{eq:backbone-annulus} for $p_B(r,R)$.

\textit{The quantum gravity approach.}---
Depending on whether $\eta$ touches $\partial \D$, the partition function $Z(L, \ell)$ of the random surface bounded 
by $\eta$ and $\partial\D$ can be  divided into two parts: $Z^{\rm nt}(L, \ell)$ and $Z^{\rm t}(L, \ell)$, where $Z^{\rm nt}(L, \ell)$ 
corresponds to when $\eta$ does not touch $\partial \D$. From the triangulation setting in Figure~\ref{fig:tri}, we see that $Z^{\rm nt}(L, \ell)$ describes a random annulus 
in the pure quantum gravity with boundary lengths $L$ and $\ell$, coupled with a percolation configuration for which the monochromatic two-arm crossing occurs. In the Liouville framework, this random annulus can be decomposed into three components: the Liouville field, the conformal matter, and the bosonic ghost field~\cite{david-conformal-gauge, dk-qg}. We conformally parameterize the annulus as the finite cylinder $\mathcal{C}_\tau$ obtained by identifying $[0,\tau] \times \{0\}$ with $[0,\tau] \times \{1\}$ on $[0,\tau]\times [0,1]$. Then the modulus $\tau$ of the annulus is random and the partition function of a given $\tau$ is the product of the partition function of three components. In our case, the partition function for the conformal matter is simply $p_B(e^{-2 \pi \tau},1)$. The ghost partition function on $\mathcal C_\tau$ is $Z_{\rm ghost}(\tau) = \eta(2 i \tau)^2$~\cite{DHokerPhong86, Martinec-annulus}, where $\eta(z) = e^{\frac{i \pi z}{12}} \prod_{n=1}^\infty (1-e^{2 n i \pi z})$ is the Dedekind eta function. Therefore for test functions $f,g$ we have $ \iint_0^\infty e^{-\nu_1 L} e^{-\nu_2 \ell} f(L) g(\ell)Z^{\rm nt}(L, \ell)d\ell dL$  
\begin{equation}\label{eq:liouville-ghost}
\propto \int_0^\infty p_B(e^{-2 \pi \tau},1)  \langle f(L_0)g(L_1) \rangle_\tau Z_{\rm ghost} (\tau) \, d\tau,
\end{equation}
where $ \langle f(L_0)g(L_1) \rangle_\tau$ is averaging over the Liouville theory on $\mathcal C_\tau$ with boundary cosmological constants $\nu_1$ and $\nu_2$, and $L_0$ and $L_1$ are the two boundary lengths.  Similarly to~\eqref{eq:QD(1,0)}, we set the bulk cosmological constant to be 0 since there is no area constraint on the surface. See Appendix~\ref{subsec:quantum-annulus} for more details on~\eqref{eq:liouville-ghost}.

As done in~\cite{ARS2022moduli}, for $\nu_1 = \nu_2 = 0$, solving the Liouville theory on the annulus gives 
\begin{equation}\label{eq:liouville-annulus}
\langle L_0 e^{-L_0} L_1^{ix} \rangle_\tau= \frac{1}{\sqrt{2} \eta(2 i \tau)} \cdot e^{-\frac{\pi \gamma^2 x^2 \tau}{4}} \cdot \frac{\pi \gamma x \Gamma(1+ix)}{2 \sinh(\frac{\gamma^2}{4} \pi x)}.\end{equation}
 (The factor $\frac{1}{\sqrt{2} \eta(2 i \tau)}$ is not present in~\cite{ARS2022moduli} due to a different normalization of the Liouville theory on $\mathcal C_\tau$.)  On the other hand, by~\eqref{eq:backbone-CR-formula} and~\eqref{eq:key},
\begin{equation}\label{eq:Z}
  \int_0^\infty Z(L,\ell)  \ell^{ix} d\ell= \frac{3 \sqrt{3}}{4} \frac{\sinh(\pi x)}{\sinh(\frac{4 \pi x}{3}) + \frac{\sqrt{3}}{2} x} L^{ix - 1}. 
\end{equation}
The surface corresponding to $ Z^{\rm t}(L,\ell) $ can be analyzed in the probabilistic framework \cite{She16a, MR4340069}, as done in~\cite{SXZ24-annulus}; see Appendix~\ref{subsec:quantum-annulus} for further background. The counterpart of~\eqref{eq:Z} is
\begin{equation}\label{eq:Zt}
\int_0^\infty Z^{\rm t}(L,\ell)  \ell^{ix} d\ell= \frac{3 \sqrt{3}}{4} \frac{\sinh(\frac{\pi x}{3})}{\sinh(\frac{2 \pi x}{3})} L^{ix - 1}.   
\end{equation}
Since $Z^{\rm nt}(L,\ell)  = Z(L,\ell) -Z^{\rm t}(L,\ell)$, combining ~\eqref{eq:liouville-ghost}--\eqref{eq:Zt}, we arrive at 
\begin{equation}\label{eq:backbone-annulus-laplace}
\begin{aligned}
&\quad \int_0^\infty p_B(e^{-2 \pi \tau},1) \eta(2 i \tau) e^{-\frac{2\pi x^2 \tau}{3}} d\tau \\
&= \frac{\sqrt{3}}{x} \bigg( \frac{\sinh(\frac{2}{3}\pi x) \sinh(\pi x)}{\sinh(\frac{4}{3} \pi x) + \frac{\sqrt{3}}{2} x} - \sinh(\frac{1}{3} \pi x) \bigg).
\end{aligned}
\end{equation}
Taking the inverse Laplace transform of~\eqref{eq:backbone-annulus-laplace} with respect to $x^2$ gives $p_B(e^{-2 \pi \tau},1) \eta(2 i \tau)$, which yields~\eqref{eq:backbone-annulus}. See Appendix~\ref{subsec:quantum-annulus} for details.

\section{Conclusions}

To summarize, we use the conformal radius of a random domain bounded by an SLE curve to encode the backbone exponent~\eqref{eq:backbone-CR}. Using 2D quantum gravity, we relate the conformal radius to the boundary length partition function of certain random surfaces~\eqref{eq:key}, which can be solved using the integrability of Liouville conformal field theory. This method is broadly applicable to percolation models whose scaling limits are described by the conformal loop ensemble (CLE). This includes the critical $Q$-Potts random cluster model~\cite{FK72} with $Q\in (0,4]$, where $Q=1$ corresponds to Bernoulli bond percolation. For general $Q$, the scaling limit is described by CLE with $\kappa = \frac{4\pi}{\pi - \arccos(\sqrt{Q}/2)}\in [4,8)$. In our derivation of~\eqref{eq:backbone-exponent} in~\cite{NQSZ-backbone}, 
we in fact treat CLE$_\kappa$ with $\kappa\in (4,8)$ uniformly. Under the scaling limit assumption, we obtained the backbone exponent for general $Q$:
\begin{equation*}
    \sin(\frac{8 \pi}{\kappa})\sqrt{\frac{\kappa  x}{2} + (1-\frac{\kappa}{4})^2} - \sin \bigg( \frac{8\pi}{\kappa}\sqrt{\frac{\kappa x}{2} + (1-\frac{\kappa}{4})^2} \bigg) = 0.
\end{equation*}
Specializing to $\kappa=6$, we get~\eqref{eq:backbone-exponent}. In~\cite{ASYZ-non-simple}, we derive the nested-path exponent defined in~\cite{STZJ22} for the random cluster model. Our method for deriving~\eqref{eq:backbone-annulus} is also  broadly applicable. We plan to derive the  analog of~\eqref{eq:backbone-annulus} for the random cluster model in a future work. 

The CFT aspect for percolation is a classical yet active topic \cite{Nienhuis-1984, DiFrancescoSaleurZuber87, cardy-formula, Nivesvivat:2023kfp}. The one-arm exponent and the polychromatic arm exponents have a CFT interpretation~\cite{He:2020rfk}.  It would be highly desirable to find a CFT interpretation for the backbone exponent $x_B$. The expansion~\eqref{eq:backbone-annulus} is reminiscent of the closed channel expansion in boundary CFT, where $\mathcal S$ can be thought of as the bulk spectrum. We can also expand in the open channel; see~\cite[Equation (1.7)]{SXZ24-annulus}. In this expansion a logarithmic structure emerges. 
We thus suspect that there is an interesting logarithmic CFT that captures $x_B$. The monochromatic $k$-arm exponent for $k \geq 3$ lies strictly between the polychromatic $k$- and $(k+1)$-arm exponents \cite{BN2011, jacobsen2002monochromatic}. Its precise value remains unknown. Unlike the backbone exponent, we have not found an effective conformal radius encoding that can be solved via the method in~\cite{AHS21}.

\appendix

\section{Supplementary Material}\label{sec:supple}

We provide further background and additional details in our derivation of~\eqref{eq:backbone-exponent} and \eqref{eq:backbone-annulus}.  
First, we elaborate on the encoding of the percolation exponents using CLE, specifically~\eqref{eq:one-arm-CR} and \eqref{eq:backbone-CR}. Next, we offer a brief overview of Liouville field theory and 2D quantum gravity and provide further details on~\eqref{eq:key0} and~\eqref{eq:key}, which ultimately lead to~\eqref{eq:backbone-CR-formula}. Finally, we focus on quantum gravity on the annulus and present the derivation details for~\eqref{eq:backbone-annulus}.

\subsection{Percolation exponents and CLE}\label{subsec:percolation}

The conformal loop ensemble (CLE)  with parameter $\kappa \in (8/3,8)$, as introduced by Sheffield~\cite{SheffieldCLE}, is a random collection of loops, each of which is an SLE$_\kappa$ curve. Given two loops in a CLE, their enclosed regions  are either nested or disjoint. For $\kappa \in (8/3, 4]$, the loops are simple and do not touch each other, whereas for $\kappa \in (4,8)$, the loops are non-simple and may touch each other. For $Q \in (0,4]$, CLE with $\kappa = \frac{4\pi}{\pi - \arccos(\sqrt{Q}/2)}\in [4,8)$ is conjectured to be the scaling limit of the percolation interfaces in the critical $Q$-Potts random cluster model~\cite{FK72}. Specifically, for any $\epsilon>0$, as the mech size tends to 0, all interfaces with diameter at least $\epsilon$ are conjectured to converge in law to the CLE loops with diameter at least $\epsilon$. In particular, this conjecture was proved for critical Bernoulli percolation on the triangular lattice~\cite{Sm01}, which corresponds to $Q=1$ and $\kappa = 6$, and for critical FK-Ising percolation on the square lattice~\cite{Smirnov-10, Kemppainen-Smirnov-19}, 
which corresponds to $Q = 2$ and $\kappa = 16/3$. 
Various percolation crossing events can be encoded by CLE loops. For example, let $\eta_0$ be the outermost CLE$_6$ loop on the unit disk that surrounds the origin, and let $d(0,\eta_0)$ denote the Euclidean distance between 0 and $\eta_0$. Recall the annulus one-arm crossing probability $p(r,1)$. Then, we have 
\begin{equation}\label{eq:one-arm}
  \mathbb{P}[d(0, \eta_0) \leq r] = p(r,1)\quad \textrm{for }0<r<1.
\end{equation}

\subsubsection{Detailed derivation of~\eqref{eq:one-arm-CR}}

Let ${\rm CR}(0, D)$ be the conformal radius as defined below Figure~\ref{fig:1}. The Koebe 1/4 theorem (see, e.g.,~\cite{Conway-complex}) states
$$
d(0, \eta_0) \leq {\rm CR}(0, D) \leq 4 d(0, \eta_0).
$$ Since the one-arm exponent $x_1$ for percolation satisfies $p(r,1) = r^{x_1 + o(1)}$, it follows from~\eqref{eq:one-arm} that
$$
\mathbb{P}[{\rm CR}(0, D) \leq r] = r^{x_1 + o(1)},
$$
which  yields~\eqref{eq:one-arm-CR}.

\subsubsection{Detailed derivation of~\eqref{eq:backbone-CR}}

The monochromatic two-arm event can be encoded using the filled percolation interfaces for the following reason. Fix $0<r<1$ and consider the subgraph formed by black vertices in the annulus $A(r,1)$. We view the inner and outer boundaries of $A(r,1)$ as two vertices, $x$ and $y$. By the vertex version of Menger's theorem (see, e.g.,~\cite{graph-theory}), the maximal number of vertex-disjoint paths from $x$ to $y$, i.e., the number of disjoint black crossings, is equal to the size of the minimum vertex cut, i.e., the minimal number of black vertices needed to remove to disconnect the inner and outer boundaries of $A(r,1)$. If the outermost filled percolation interface has distance less than $r$ to the origin, then we need to remove at least two black vertices to disconnect the inner and outer boundaries of $A(r,1)$, and thus, there exist two disjoint black crossings. In the continuum limit, the filled percolation interfaces converge to the outer boundaries of CLE$_6$ loops. These outer boundaries are simple loops that locally look like SLE$_{8/3}$ curves~\cite{LSW_CR_chordal}, while different loops may touch each other. Let $\eta$ be the outermost such simple loop that surrounds 0. Then, the previous discrete argument implies
\begin{equation}\label{eq:backbone-detail}
\mathbb{P}[d(0, \eta) \leq r] \leq  p_B(r,1) \quad \mbox{for $0<r<1$}.
\end{equation}

\begin{figure}[h]
\vspace{-0.2cm}
\begin{center}
\includegraphics[angle=0,width=.93290\linewidth]{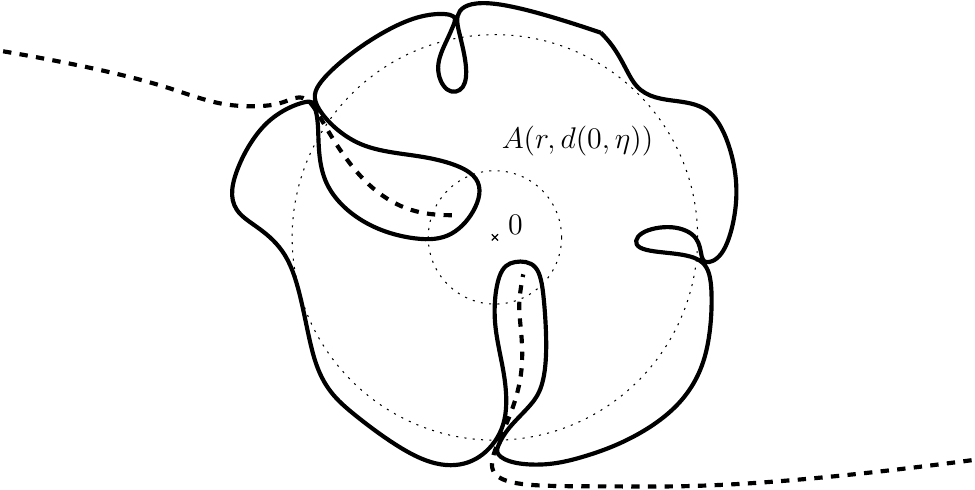}
\end{center}
\vspace{-0.5cm}\caption{The two dashed lines represent two disjoint black crossings of $A(r, 1)$ on the event $d(0,\eta)>r$.}\label{fig:supp1}
\vspace{-0.3cm}
\end{figure}

Let $\widetilde x_B$ be the exponent such that $\mathbb{P}[d(0, \eta) \leq r] = r^{\widetilde x_B+o(1)}$. Then, \eqref{eq:backbone-detail} implies that $\widetilde x_B \geq x_B$. We now elaborate that $\widetilde x_B \leq x_B$, and thus the two probabilities in~\eqref{eq:backbone-detail} share the same exponent as $r \to 0$. If $d(0,\eta)>r$, there may still exist two disjoint black crossings of $A(r,1)$; see Figure~\ref{fig:supp1}. However, in this case, there must be two white crossings separating the two black crossings in the annulus $A(r, d(0,\eta))$, and thus, an alternating four-arm event occurs. The probability of this event is known to be $(\frac{r}{d(0,\eta)})^{x_4+o(1)}$, where $x_4 = \frac{5}{4}$~\cite{ADA99}. By partitioning the values of $d(0,\eta)$ into dyadic intervals $(2^n r, 2^{n+1}r]$ for $0 \leq n \leq \lfloor \log_2 r^{-1} \rfloor $, we obtain
\begin{align*}
&p_B(r,1) \leq \mathbb{P}[d(0,\eta) \leq r] \\
&\quad + \sum_{n=0}^{ \lfloor \log_2 r^{-1} \rfloor } \mathbb{P}[2^n r < d(0,\eta) \leq 2^{n+1} r] \times (2^{-n})^{x_4+o(1)}.
\end{align*}
Using $\mathbb{P}[d(0, \eta) \leq r] = r^{\widetilde x_B+o(1)}$ and $\widetilde x_B < x_4$, we obtain
\begin{align*}
    p_B(r,1) \leq r^{\widetilde x_B+o(1)}.
\end{align*}
Therefore, the two probabilities in~\eqref{eq:backbone-detail} share the same exponent as $r \to 0$. In particular,
$$
\mathbb{P}[d(0, \eta) \leq r] = r^{x_B+o(1)}.
$$
Combining this with the Koebe 1/4 theorem $d(0, \eta) \leq {\rm CR}(0, D_b) \leq 4 d(0, \eta)$ yields~\eqref{eq:backbone-CR}, similarly to~\eqref{eq:one-arm-CR}.

\subsection{Liouville field theory and 2D quantum gravity}\label{subsec:liouville}

We begin with an overview of Liouville field theory and its connection to 2D quantum gravity. Next, we review the coupling theory between SLE and Liouville quantum gravity, from which we derive~\eqref{eq:key0}. We then explain the derivation of the key equation~\eqref{eq:key}.

\subsubsection{Liouville field theory}

The correlation function of Liouville field theory is
\begin{equation}\label{eq:def-correlation}
\int_{\phi: \Sigma \to \mathbb{R}} \prod_{i=1}^n e^{\alpha_i \phi(x_i)} e^{-S_L[\phi]} D\phi,
\end{equation}
where $\Sigma$ is a 2D Riemannian manifold, $\alpha_i \in \mathbb{R}$, and $x_i$ are marked points on $\Sigma$. Here, $S_L[\phi]$ is the Liouville action on $\Sigma$.
We need the cases where $\Sigma$ is the unit disk $\D$ or the cylinder $\mathcal{C}_\tau$ obtained by identifying $[0,\tau] \times \{0\}$ with $[0,\tau] \times \{1\}$ on $[0,\tau]\times [0,1]$. The Liouville action on $\D$ is provided in~\eqref{eq:liouville-action}. For $\mathcal{C}_\tau$, the action $S_L[\phi]$ is given by
\begin{equation}\label{eq:action-annulus}
\int_{\mathcal C_\tau} (\frac{1}{4\pi}|\nabla \phi|^2 + \mu e^{\gamma \phi}) d^2x + \int_{\partial_1 \mathcal C_\tau} \nu_1 e^{\frac{\gamma}{2} \phi} d l +  \int_{\partial_2 \mathcal C_\tau} \nu_2 e^{\frac{\gamma}{2} \phi} d l
\end{equation}
where $\partial_1 \mathcal C_\tau$ and $\partial_2 \mathcal C_\tau$ are the two boundaries of $\mathcal C_\tau$, and $\mu, \nu_1,\nu_2$ are cosmological constants. The coupling constant $\gamma \in (0,2)$ determines the background charge $Q = \frac{2}{\gamma} + \frac{\gamma}{2}$ and the central charge $c_L = 1 + 6Q^2$. Liouville correlation functions can be explicitly solved under the conformal field theory framework~\cite{BPZ84}, 
which is done in physics by~\cite{DO94, ZZ96, PT02} and rigorously in math in~\cite{KRV20, GKRV20bootstrap}; see~\cite{GKR-review} for a review.
In our derivation of~\eqref{eq:backbone-CR-formula} and~\eqref{eq:backbone-annulus}, we need the boundary structure constants on the disk solved in~\cite{RZ22} and the annulus partition function solved in~\cite{Wu-annulus}, in the case where the bulk cosmological constant is zero.

\subsubsection{2D quantum gravity and random triangulation}

Random triangulation is a discrete model for 2D quantum gravity, where the uniform case corresponds to pure gravity; see~\cite{LeGall-ICM, Sheffield-2022ICM} for a mathematical overview. 
Taking a uniform random triangulation of the sphere with $n$ faces and sending $n$ to infinity, we obtain a random surface with unit area. It is natural to define a surface with unrestricted area by reweighting the area with a random variable $A$ sampled from the infinite measure $A^{-7/2} dA$, since the total number of triangulations with $n$ faces grows like $\alpha^n n^{-7/2}$ for some $\alpha\in (0,\infty)$~\cite{Tutte63, Brown64}. 
Similarly, we can use the random triangulation of the disk to define a random disk with one interior marked point and no area or boundary length constraints. The total number of random triangulations of the disk with one interior marked point, $n$ faces, and perimeter $p$ grows like $\alpha^n \beta^p n^{-3/2} p^{-1/2} e^{-C p^2/n}$ for $\alpha$ above and some $\beta\in (0,\infty)$~\cite{LQG-disk}. If we assign the weight $\alpha^{-n} \beta^{-p}$ (known as the critical Boltzmann weight) and sum over $n$, we see that the total weight grows like $p^{-3/2}$ as $p \to \infty$. Therefore, the boundary length of the resulting random disk follows a law proportional to $L^{-3/2} dL$.

2D pure quantum gravity can alternatively be described by Liouville field theory with $c_L=26$ and $\gamma = \sqrt{8/3}$, where the exponential of field defines a random geometry that can be viewed as the gravitational background. To understand this geometric construction, note that when the bulk and boundary cosmological constants are set to 0, the field $\phi$ sampled from the distribution $e^{-S_L[\phi]} D\phi$ on $\Sigma$ is simply the Gaussian free field (GFF), which is a random generalized function. Using a normalization procedure known as Gaussian multiplicative chaos theory, we can make sense of the area measure $e^{\gamma \phi} d^2x$ on $\Sigma$ and the boundary length measure $e^{\frac{\gamma}{2} \phi} dl$ on $\partial \Sigma$~\cite{DS-KPZ-invent}. The potential terms $\int_\Sigma e^{\gamma \phi} d^2 x $ and $ \int_{\partial \Sigma} e^{\frac{\gamma}{2} \phi} dl $ in the action $S_L[\phi]$ correspond to the total area and boundary length of this random surface. Therefore, Liouville field theory can be understood as reweighting the GFF by the total area and boundary length. Moreover, by Girsanov's theorem, vertex operators $e^{\alpha_i \phi(x_i)}$ correspond to adding the singularity $-\alpha_i \log|\cdot-x_i|$ to the GFF.

For the sphere case, sample $\phi$ from $e^{-S_L[\phi]} D\phi$ with bulk cosmological constant 0. Then, the resulting random sphere describes the scaling limit of the random triangulation of the sphere with no area constraint. In this case, the law of the total area measure of the Liouville field is proportional to $A^{-1-2Q/\gamma} dA$. Now $\gamma = \sqrt{8/3}$ and $Q = \sqrt{25/6}$ give $1+2Q/\gamma = 7/2$, which is consistent with the exponent obtained from triangulation enumeration.
For the disk case, sample $\phi$ from $e^{\gamma \phi(0)} e^{-S_L[\phi]} D\phi$ on $\D$ with all cosmological constants 0. Then, the resulting random disk describes the scaling limit of the random triangulation of the disk with one interior marked point and no area or boundary length constraints. In particular, the law of the total boundary length of the Liouville field is proportional to $L^{-3/2} dL$, which again matches the triangulation result. 
From both perspectives, we see that for $\QD_1(L)$ defined below~\eqref{eq:QD(1,0)}, its partition function $|{\rm QD}_1(L)| $ of  is $CL^{-\frac32}$ for some constant $C>0$. 

\subsubsection{Derivation of~\eqref{eq:key0}: cut and glue random surfaces}

Sample a critical Bernoulli site percolation on top of a random triangulation of the disk as in Figure~\ref{fig:tri}. 
A crucial assumption in 2D quantum gravity is that the continuum limit of percolation is still CLE$_6$, independently of the geometric background. The random geometry affects only the lengths of the loops.  A version of this result was rigorously proved in~\cite{HS19}.  
Bernoulli percolation on random triangulations has a greater degree of spatial independence compared to its regular lattice counterpart. 
A specific example is shown in Figure~\ref{fig:tri}: the outermost filled percolation interface divides the random triangulation into two independent surfaces conditioned on the number of vertices along the filled interface. For $L>0$, we write $Z_{\rm whole}(L)$ as the partition function of the original random triangulation with boundary length $L$. Then we have 
\begin{equation}\label{eq:discrete-partition}
    Z_{\rm whole}(L) = \sum_{\ell} Z_{\rm outside}(L,\ell) \times \ell \times Z_{\rm inside}(\ell),
\end{equation}
where $Z_{\rm inside}, Z_{\rm outside}$ are Boltzmann-weighted partition functions of the two random triangulations separated by the  outermost filled percolation interface, whose length is denoted by $\ell$. The additional factor $\ell$ on the right-hand side counts the number of ways the two surfaces can be glued together to recover the original disk, since there is no marked  point on their common boundary.

Equation~\eqref{eq:key0} is the exact continuum analog of~\eqref{eq:discrete-partition}. In fact,  there is a way to derive such equations directly in the continuum using the quantum zipper method pioneered by Sheffield~\cite{She16a}. 
Using this method, we proved in~\cite{SXZ24-annulus} that similar spatial independence holds when we use the curve $\eta$ to cut the random disk of length $L$. Furthermore,  the quantum zipper method shows that the law of the random disk inside  $\eta$ is an independent copy of the original random disk given its boundary length. Namely, the continuum analogs of $Z_{\rm whole}(L)$ and $Z_{\rm inside}(\ell)$ take the same form, which are $L^{-3/2}$ and $\ell^{-3/2}$ respectively.
(Note that the discrete counterpart of this statement is not exactly true, since the surface bounded by the outermost filled percolation interface is a random triangulation of the disk decorated with a percolation configuration subject to complicated constraints.) Writing $Z(L,\ell)$ as the continuum analog of $Z_{\rm outside}(L,\ell)$, we get~\eqref{eq:key0}.

\subsubsection{Derivation of~\eqref{eq:key}: change of vertex insertion}

Recall that the random disk ${\rm QD}_1$ can be described by the Liouville field sampled from $e^{\gamma \phi(0)} e^{-S_L[\phi]} D\phi$ on $\D$. For general $\alpha \in \mathbb{R}$ and a field $\phi$ sampled from $e^{\alpha \phi(0)} e^{-S_L[\phi]} D\phi$ on $\D$, the law of the boundary length is proportional to $L^{\frac{2}{\gamma}(\alpha - \gamma) -\frac{3}{2}}$. To obtain~\eqref{eq:key}, we change the vertex operator at the origin from $e^{\gamma \phi(0)}$ to $e^{\alpha \phi(0)}$ on both sides of~\eqref{eq:key0} while keeping the random surface with partition function $Z(L,\ell)$ unchanged. Take $\alpha = \gamma + \frac{\gamma}{2} a$ such that $\frac{2}{\gamma}(\alpha - \gamma) -\frac{3}{2} = -\frac{3}{2} + a$. This gives the term $L^{ -\frac{3}{2} + a}$ on the right-hand side of~\eqref{eq:key}.

Since the random disk bounded by $\eta$ is also a sample from  ${\rm QD}_1$, changing the insertion from $e^{\gamma\phi(0)} $ to $e^{\alpha \phi(0)}$ results in the same shift of the partition function from $\ell^{ -\frac{3}{2}}$ to $\ell^{ -\frac{3}{2} + a}$, if the random disk were parameterized by $\D$ instead of the domain $D_b$ bounded by $\eta$. This change of domain results in an additional conformal factor, which is exactly ${\rm CR}(0, D_b)^{x(a)}$ on the right-hand side of~\eqref{eq:key}. To see the effect of the coordinate change, 
for a simply connected domain $D$ containing $0$, let $\langle e^{\alpha \phi(0)} \rangle_D$ be the correlation function $ \int_{\phi: D \to \mathbb{R}} e^{\alpha \phi(0)} e^{-S_L[\phi]} D\phi$. For a conformal map $f:(\D, 0) \to (D_b,0)$, we have
$$
    \langle e^{\alpha \phi(0)} \rangle_{D_b} = |f'(0)|^{-2\Delta_\alpha} \langle e^{\alpha \phi(0)} \rangle_\D,
$$
where $\Delta_\alpha = \frac{\alpha}{2}(Q - \frac{\alpha}{2})$ is the conformal dimension of $e^{\alpha\phi(0)}$. Since by definition ${\rm CR}(0, D_b) = |f'(0)|$, the additional factor is given by
$$
\frac{\langle e^{\alpha \phi(0)} \rangle_\D}{\langle e^{\alpha \phi(0)} \rangle_{D_b}} \Big{/} \frac{\langle e^{\gamma \phi(0)} \rangle_\D}{\langle e^{\gamma \phi(0)} \rangle_{D_b}} = {\rm CR}(0, D_b)^{2\Delta_\alpha - 2 \Delta_\gamma}.
$$
When $\alpha = \gamma + \frac{\gamma}{2} a$, we have $2\Delta_\alpha - 2 \Delta_\gamma = -\frac{1}{3}a(a-1) =  x(a)$. This yields~\eqref{eq:key}.

\subsection{Quantum gravity on the annulus}\label{subsec:quantum-annulus}

Recall that $Z(L, \ell)$ is the continuum analog of $Z_{\rm outside} (L,\ell)$. In the discrete, $Z_{\rm outside} (L, \ell)$ can be divided into two parts $Z_{\rm outside}^{\rm nt}$ and $Z_{\rm outside}^{\rm t}$ depending on whether the outermost filled interface as in Figure~\ref{fig:tri} touches the boundary or not (in Figure~\ref{fig:tri} it does not touch). The continuum analogs of these are respectively $Z^{\rm nt}$ and $Z^{\rm t}$. In fact, both of these random surfaces can be described by Liouville field theory, which yields~\eqref{eq:liouville-ghost} and~\eqref{eq:Zt}, respectively.

We first consider $Z^{\rm nt}$. Recall that $Z_{\rm outside}^{\rm nt}$ is the Boltzmann-weighted partition function of a random triangulation of an annulus decorated with a percolation configuration subject to the monochromatic two-arm event. Without the percolation decoration, this random triangulation is the discrete model for pure gravity on the annulus. 
Its Liouville description is well known in the bosonic string theory~\cite{polyakov1981quantum}, which is 
\begin{equation*}
    Z_{\rm ghost}(\tau) Z_{\rm Liouville}(\tau) d\tau. 
\end{equation*}
Here, $Z_{\rm ghost}(\tau) $ is the partition function for the ghost CFT with central charge $-26$, coming from the Faddeev-Popov determinant for the conformal gauge-fixing. For the annulus with modulus $\tau$, it is given in~\cite{DHokerPhong86, Martinec-annulus} that $Z_{\rm ghost}(\tau)=\eta(2i\tau)^2$.
The Liouville partition function  $Z_{\rm Liouville}(\tau) $ is given by the Liouville action $S_L[\phi]$~\eqref{eq:action-annulus} on the annulus:
$Z_{\rm Liouville}(\tau) = \int_{\phi: \mathcal{C}_\tau \to \mathbb{R}} e^{-S_L[\phi]} D\phi$. 

Now that $Z_{\rm outside}^{\rm nt}$ is the random triangulation decorated with a percolation configuration where  the monochromatic two-arm event occurs, we can view the decorating percolation as a conformal matter with central charge 0 and partition function $p_B(e^{-2 \pi \tau}, 1)$.
For gravity coupled to conformal matter on the annulus, the Liouville description becomes
\begin{equation*}
Z_{\rm ghost}(\tau) Z_{\rm Liouville}(\tau) Z_{\rm matter}(\tau) d\tau.
\end{equation*}
Applying this to $Z^{\rm nt}$, we obtain~\eqref{eq:liouville-ghost}.

The random surface corresponding to $Z^{\rm t}$ is quite intriguing. In fact, when the curve $\eta$ touches $\partial \D$, it will touch at infinitely many points. The surface bounded by $\eta$ and $\partial \D$ consists of a countable collection of topological disks. Again based on the quantum zipper method, it was proved in~\cite{SXZ24-annulus} that the law of this collection of random disks is Poissonian, and conditioning on their boundary lengths, these disks are independent copies of random disks that can be described by Liouville theory on the disk with two boundary insertions. This allows \cite{SXZ24-annulus} to derive the exact expression~\eqref{eq:Zt} for $Z^{\rm t}$.

\subsubsection{Derivations of~\eqref{eq:backbone-annulus-laplace} and~\eqref{eq:backbone-annulus}}
We first provide further background on~\eqref{eq:liouville-annulus}. Using the CFT framework~\cite{BPZ84}, the Liouville annulus partition function was solved in~\cite{Wu-annulus}. If we set the bulk cosmological constant to 0 and the two boundary cosmological constants to $\nu_1,\nu_2$, then the partition function becomes
$\langle e^{-\nu_1 L_0} e^{-\nu_2 L_1} \rangle_\tau$ where the average is over a Liouville field theory on $\mathcal{C}_\tau$ with all cosmological constants 0, and $L_0$ and $L_1$ are the two boundary lengths. Based on this expression, equation~\eqref{eq:liouville-annulus} was derived in~\cite{ARS2022moduli}.

By~\eqref{eq:key} and~\eqref{eq:backbone-CR-formula}, and taking $b = -\frac{1}{2}+a$, we have
$$
\int_0^\infty Z(L, \ell) \ell^b d \ell = \frac{3 \sqrt{3}}{4} \frac{\sin(\pi b)}{\sin(\frac{4 \pi b}{3}) + \frac{\sqrt{3}}{2} b} L^{b-1}.
$$
Analytically extending both sides of the equation and setting $b = ix$ yields~\eqref{eq:Z}. Next, we derive~\eqref{eq:backbone-annulus-laplace}. On the one hand, by~\eqref{eq:Z}, \eqref{eq:Zt}, and $Z^{\rm nt}(L,\ell) = Z(L,\ell) - Z^{\rm t}(L,\ell)$, we have
\begin{align*}
    &\quad \iint_0^\infty Le^{-L} Z^{\rm nt}(L,\ell) \ell^{ix} dL d\ell\\
    &=\int_0^\infty L e^{-L}\bigg( \frac{3 \sqrt{3}}{4} \frac{\sinh(\pi x)}{\sinh(\frac{4 \pi x}{3}) + \frac{\sqrt{3}}{2} x}  - \frac{3 \sqrt{3}}{4} \frac{\sinh(\frac{\pi x}{3})}{\sinh(\frac{2 \pi x}{3})} \bigg)\\
    &\qquad \quad \times L^{ix-1} dL\\
    &= \frac{3 \sqrt{3}}{4} \frac{\Gamma(ix+1)}{\sinh(\frac{2 \pi x}{3})} \bigg( \frac{\sinh(\frac{2}{3}\pi x) \sinh(\pi x)}{\sinh(\frac{4}{3} \pi x) + \frac{\sqrt{3}}{2} x} - \sinh(\frac{1}{3} \pi x) \bigg).
\end{align*}
On the other hand, by~\eqref{eq:liouville-ghost} and~\eqref{eq:liouville-annulus},
\begin{align*}
    &\quad \iint_0^\infty Le^{-L} Z^{\rm nt}(L,\ell) \ell^{ix} d\ell\\
    &\propto \int_0^\infty p_B(e^{-2 \pi \tau},1) \langle L_0 e^{-L_0} L_1^{ix} \rangle_\tau Z_{\rm ghost}(\tau) d\tau \\
    &= \int_0^\infty p_B(e^{-2 \pi \tau},1)\frac{1}{\sqrt{2} \eta(2 i \tau)} \cdot e^{-\frac{\pi \gamma^2 x^2 \tau}{4}} \cdot \frac{\pi \gamma x \Gamma(1+ix)}{2 \sinh(\frac{\gamma^2}{4} \pi x)} \\
    &\qquad \quad \times \eta(2 i \tau)^2 d\tau \\
    &= \frac{\pi x \Gamma(ix+1)}{\sqrt{3} \sinh(\frac{2 \pi x}{3})} \int_0^\infty p_B(e^{-2 \pi \tau},1) \eta(2 i \tau) e^{-\frac{2\pi x^2 \tau}{3}} d\tau.
\end{align*}
Combining the above two equations, we obtain~\eqref{eq:backbone-annulus-laplace} up to a constant. This constant can be fixed using the condition $\lim_{\tau \to 0} p_B(e^{-2 \pi \tau},1) = 1$, which gives~\eqref{eq:backbone-annulus-laplace}. 

Finally, we derive~\eqref{eq:backbone-annulus}. Replacing $\frac{2\pi x^2}{3}$ in~\eqref{eq:backbone-annulus-laplace} with $t$, we obtain that $\int_0^\infty e^{-t\tau} p_B(e^{-2 \pi \tau},1) \eta(2 i \tau) d\tau = g(t)$ for any $t \in (0, \infty)$, where $g(t)$ is given by
    \begin{equation*}
    g(t) = \frac{\sqrt{3}}{x} \bigg( \frac{\sinh(\frac{2}{3}\pi x) \sinh(\pi x)}{\sinh(\frac{4}{3} \pi x) + \frac{\sqrt{3}}{2} x} - \sinh(\frac{1}{3} \pi x) \bigg)
    \end{equation*}
    with $x = \sqrt{\frac{3 t}{2 \pi}}$.
    By the inverse Laplace transform,
    $$
    p_B(e^{-2 \pi \tau},1) \eta(2 i \tau) = \frac{1}{2 \pi i} \int_{-i \infty}^{i \infty} e^{\tau t} g(t) dt.
    $$
    Next, we deform the integral contour and apply Cauchy's residue theorem which yields
    \begin{equation}\label{eq:backbone-annulus-0}
    p_B(e^{-2 \pi \tau},1) \eta(2 i \tau) = \sum_{s \in \mathcal{S}'} {\rm Res}(e^{\tau t} g(t), s)
    \end{equation}
    where $\mathcal{S}'$ is the set of poles of $g(t)$ in $\mathbb{C}$, and ${\rm Res}(e^{\tau t} g(t), s)$ is the residue of $e^{\tau t} g(t)$ at $t=s$. Specifically, $\mathcal{S}'$ consists of all the complex solutions to $\sinh(4 \pi \sqrt{\frac{t}{6\pi}}) + \frac{3}{2}\sqrt{\frac{t}{2\pi}} = 0$ except $0$ and $-\frac{2 \pi}{3}$. After simplification, \eqref{eq:backbone-annulus-0} reduces to~\eqref{eq:backbone-annulus}.

\medskip
\noindent\textbf{Acknowledgements.} P.N.\ is partially supported by a GRF grant from the Research Grants Council of the Hong Kong SAR (project CityU11318422).  W.Q. and X.S. are supported by National Key R\&D Program of China (No.\ 2023YFA1010700). Z.Z.\ is partially supported by NSF grant DMS-1953848. We thank B. Duplantier and two anonymous referees for helpful comments.

\bibliographystyle{alpha}
\bibliography{ref-math}

\newcommand{\etalchar}[1]{$^{#1}$}
\begin{thebibliography}{GKRV24}

\bibitem[ADA99]{ADA99}
Michael Aizenman, Bertrand Duplantier, and Amnon Aharony.
\newblock Path-crossing exponents and the external perimeter in 2{D} percolation.
\newblock {\em Phys. Rev. Lett.}, 83:1359--1362, 1999.

\bibitem[AHS24]{AHS21}
Morris Ang, Nina Holden, and Xin Sun.
\newblock Integrability of {SLE} via conformal welding of random surfaces.
\newblock {\em Comm. Pure Appl. Math.}, 77(5):2651--2707, 2024.

\bibitem[ARS22]{ARS2022moduli}
Morris {Ang}, Guillaume {Remy}, and Xin {Sun}.
\newblock {The moduli of annuli in random conformal geometry}.
\newblock {\em arXiv e-prints}, page arXiv:2203.12398, March 2022.

\bibitem[AS03]{MR2013797}
Omer Angel and Oded Schramm.
\newblock Uniform infinite planar triangulations.
\newblock {\em Comm. Math. Phys.}, 241(2-3):191--213, 2003.

\bibitem[ASYZ24]{ASYZ-non-simple}
Morris {Ang}, Xin {Sun}, Pu~{Yu}, and Zijie {Zhuang}.
\newblock {Boundary touching probability and nested-path exponent for non-simple CLE}.
\newblock {\em arXiv e-prints}, page arXiv:2401.15904, January 2024.

\bibitem[AW95]{AMBJORN1995129}
J.~Ambjørn and Y.~Watabiki.
\newblock Scaling in quantum gravity.
\newblock {\em Nuclear Physics B}, 445(1):129--142, 1995.

\bibitem[BH57]{BH57}
S.~R. Broadbent and J.~M. Hammersley.
\newblock Percolation processes. {I}. {C}rystals and mazes.
\newblock {\em Proc. Cambridge Philos. Soc.}, 53:629--641, 1957.

\bibitem[BK87]{BOULATOV1987379}
D.V. Boulatov and V.A. Kazakov.
\newblock The ising model on a random planar lattice: The structure of the phase transition and the exact critical exponents.
\newblock {\em Physics Letters B}, 186(3):379--384, 1987.

\bibitem[BM17]{BM17-brownian-disk}
J\'er\'emie Bettinelli and Gr\'egory Miermont.
\newblock Compact {B}rownian surfaces {I}: {B}rownian disks.
\newblock {\em Probab. Theory Related Fields}, 167(3-4):555--614, 2017.

\bibitem[BN11]{BN2011}
Vincent Beffara and Pierre Nolin.
\newblock On monochromatic arm exponents for 2{D} critical percolation.
\newblock {\em Ann. Probab.}, 39(4):1286--1304, 2011.

\bibitem[BPZ84]{BPZ84}
A.~A. Belavin, A.~M. Polyakov, and A.~B. Zamolodchikov.
\newblock Infinite conformal symmetry in two-dimensional quantum field theory.
\newblock {\em Nuclear Phys. B}, 241(2):333--380, 1984.

\bibitem[Bro64]{Brown64}
William~G. Brown.
\newblock Enumeration of triangulations of the disk.
\newblock {\em Proc. London Math. Soc. (3)}, 14:746--768, 1964.

\bibitem[Car92]{cardy-formula}
John~L. Cardy.
\newblock Critical percolation in finite geometries.
\newblock {\em J. Phys. A}, 25(4):201--206, 1992.

\bibitem[Car02]{Car02}
John Cardy.
\newblock Crossing formulae for critical percolation in an annulus.
\newblock {\em J. Phys. A}, 35(41):L565--L572, 2002.

\bibitem[Car06]{Car06}
John Cardy.
\newblock The {${\rm O}(n)$} model on the annulus.
\newblock {\em J. Stat. Phys.}, 125(1):1--21, 2006.

\bibitem[CN06]{CN06-full}
Federico Camia and Charles~M. Newman.
\newblock Two-dimensional critical percolation: the full scaling limit.
\newblock {\em Comm. Math. Phys.}, 268(1):1--38, 2006.

\bibitem[Con95]{Conway-complex}
John~B. Conway.
\newblock {\em Functions of one complex variable. {II}}, volume 159 of {\em Graduate Texts in Mathematics}.
\newblock Springer-Verlag, New York, 1995.

\bibitem[Dav88]{david-conformal-gauge}
F.~David.
\newblock Conformal field theories coupled to {2-D} gravity in the conformal gauge.
\newblock {\em {M}od. {P}hys. {L}ett. {A}}, 3(17):1651--1656, 1988.

\bibitem[DBN04]{DBN04}
Youjin Deng, Henk W.~J. Bl\"ote, and Bernard Nienhuis.
\newblock Backbone exponents of the two-dimensional q-state potts model: A monte carlo investigation.
\newblock {\em Phys. Rev. E}, 69:026114, Feb 2004.

\bibitem[DFSZ87]{DiFrancescoSaleurZuber87}
P.~Di~Francesco, H.~Saleur, and J.~B. Zuber.
\newblock {Modular Invariance in Nonminimal Two-dimensional Conformal Theories}.
\newblock {\em Nucl. Phys. B}, 285:454, 1987.

\bibitem[Die17]{graph-theory}
Reinhard Diestel.
\newblock {\em Graph theory}, volume 173 of {\em Graduate Texts in Mathematics}.
\newblock Springer, Berlin, fifth edition, 2017.

\bibitem[DK89]{dk-qg}
J.~Distler and H.~Kawai.
\newblock Conformal field theory and {2D} quantum gravity.
\newblock {\em {N}ucl.{P}hys. {B}}, 321(2):509–527, 1989.

\bibitem[DMS21]{MR4340069}
Bertrand Duplantier, Jason Miller, and Scott Sheffield.
\newblock Liouville quantum gravity as a mating of trees.
\newblock {\em Ast\'erisque}, (427):viii+257, 2021.

\bibitem[dN79]{denNijs1979}
M~P~M den Nijs.
\newblock A relation between the temperature exponents of the eight-vertex and q-state potts model.
\newblock {\em Journal of Physics A: Mathematical and General}, 12(10):1857, oct 1979.

\bibitem[DO94]{DO94}
H.~Dorn and H.-J. Otto.
\newblock Two- and three-point functions in {L}iouville theory.
\newblock {\em Nuclear Phys. B}, 429(2):375--388, 1994.

\bibitem[DP86]{DHokerPhong86}
Eric D'Hoker and D.~H. Phong.
\newblock {Multiloop Amplitudes for the Bosonic Polyakov String}.
\newblock {\em Nucl. Phys. B}, 269:205--234, 1986.

\bibitem[DS11]{DS-KPZ-invent}
Bertrand Duplantier and Scott Sheffield.
\newblock Liouville quantum gravity and {KPZ}.
\newblock {\em Invent. Math.}, 185(2):333--393, 2011.

\bibitem[Dup98]{MR1666816}
Bertrand Duplantier.
\newblock Random walks and quantum gravity in two dimensions.
\newblock {\em Phys. Rev. Lett.}, 81(25):5489--5492, 1998.

\bibitem[FK72]{FK72}
C.~M. Fortuin and P.~W. Kasteleyn.
\newblock On the random-cluster model. {I}. {I}ntroduction and relation to other models.
\newblock {\em Physica}, 57:536--564, 1972.

\bibitem[FKZD22]{FKZD22}
Sheng Fang, Da~Ke, Wei Zhong, and Youjin Deng.
\newblock Backbone and shortest-path exponents of the two-dimensional $q$-state potts model.
\newblock {\em Phys. Rev. E}, 105:044122, Apr 2022.

\bibitem[GKR24]{GKR-review}
Colin {Guillarmou}, Antti {Kupiainen}, and R{\'e}mi {Rhodes}.
\newblock {Review on the probabilistic construction and Conformal bootstrap in Liouville Theory}.
\newblock {\em arXiv e-prints}, page arXiv:2403.12780, March 2024.

\bibitem[GKRV24]{GKRV20bootstrap}
Colin Guillarmou, Antti Kupiainen, R\'emi Rhodes, and Vincent Vargas.
\newblock Conformal bootstrap in {L}iouville theory.
\newblock {\em Acta Math.}, 233(1):33--194, 2024.

\bibitem[Gra99]{G99}
Peter Grassberger.
\newblock Conductivity exponent and backbone dimension in 2-d percolation.
\newblock {\em Physica A: Statistical Mechanics and its Applications}, 262(3):251--263, 1999.

\bibitem[HJS20]{He:2020rfk}
Yifei He, Jesper~Lykke Jacobsen, and Hubert Saleur.
\newblock {Geometrical four-point functions in the two-dimensional critical $Q$-state Potts model: The interchiral conformal bootstrap}.
\newblock {\em JHEP}, 12:019, 2020.

\bibitem[HRV18]{LQG-disk}
Yichao Huang, R\'emi Rhodes, and Vincent Vargas.
\newblock Liouville quantum gravity on the unit disk.
\newblock {\em Ann. Inst. Henri Poincar\'e{} Probab. Stat.}, 54(3):1694--1730, 2018.

\bibitem[HS23]{HS19}
Nina Holden and Xin Sun.
\newblock Convergence of uniform triangulations under the {C}ardy embedding.
\newblock {\em Acta Math.}, 230(1):93--203, 2023.

\bibitem[JZJ02]{jacobsen2002monochromatic}
Jesper~Lykke Jacobsen and Paul Zinn-Justin.
\newblock Monochromatic path crossing exponents and graph connectivity in two-dimensional percolation.
\newblock {\em Physical Review E}, 66(5):055102, 2002.

\bibitem[KPZ88]{KPZrelation}
V.~G. Knizhnik, Alexander~M. Polyakov, and A.~B. Zamolodchikov.
\newblock {Fractal Structure of 2D Quantum Gravity}.
\newblock {\em Mod. Phys. Lett. A}, 3:819, 1988.

\bibitem[KRV20]{KRV20}
Antti Kupiainen, R\'emi Rhodes, and Vincent Vargas.
\newblock Integrability of {L}iouville theory: proof of the {DOZZ} formula.
\newblock {\em Ann. of Math. (2)}, 191(1):81--166, 2020.

\bibitem[KS19]{Kemppainen-Smirnov-19}
Antti Kemppainen and Stanislav Smirnov.
\newblock Conformal invariance of boundary touching loops of {FK} {I}sing model.
\newblock {\em Comm. Math. Phys.}, 369(1):49--98, 2019.

\bibitem[LG14]{LeGall-ICM}
Jean-Fran\c{c}ois Le~Gall.
\newblock Random geometry on the sphere.
\newblock In {\em Proceedings of the {I}nternational {C}ongress of {M}athematicians---{S}eoul 2014. {V}ol. 1}, pages 421--442. Kyung Moon Sa, Seoul, 2014.

\bibitem[LSW01]{MR1879851}
Gregory~F. Lawler, Oded Schramm, and Wendelin Werner.
\newblock Values of {B}rownian intersection exponents. {II}. {P}lane exponents.
\newblock {\em Acta Math.}, 187(2):275--308, 2001.

\bibitem[LSW02a]{LSW-III}
Gregory~F. Lawler, Oded Schramm, and Wendelin Werner.
\newblock Analyticity of intersection exponents for planar {B}rownian motion.
\newblock {\em Acta Math.}, 189(2):179--201, 2002.

\bibitem[LSW02b]{LSW02}
Gregory~F. Lawler, Oded Schramm, and Wendelin Werner.
\newblock One-arm exponent for critical 2{D} percolation.
\newblock {\em Electron. J. Probab.}, 7:no. 2, 13, 2002.

\bibitem[LSW03]{LSW_CR_chordal}
Gregory Lawler, Oded Schramm, and Wendelin Werner.
\newblock Conformal restriction: the chordal case.
\newblock {\em J. Amer. Math. Soc.}, 16(4):917--955, 2003.

\bibitem[Mar03]{Martinec-annulus}
Emil~J Martinec.
\newblock The annular report on non-critical string theory.
\newblock {\em arXiv preprint hep-th/0305148}, 2003.

\bibitem[MSW17]{MSW2017}
Jason Miller, Scott Sheffield, and Wendelin Werner.
\newblock C{LE} percolations.
\newblock {\em Forum Math. Pi}, 5:e4, 102, 2017.

\bibitem[Nie84]{Nienhuis-1984}
Bernard Nienhuis.
\newblock {Critical behavior of two-dimensional spin models and charge asymmetry in the Coulomb gas}.
\newblock {\em J. Statist. Phys.}, 34:731--761, 1984.

\bibitem[NQSZ23]{NQSZ-backbone}
Pierre {Nolin}, Wei {Qian}, Xin {Sun}, and Zijie {Zhuang}.
\newblock {Backbone exponent for two-dimensional percolation}.
\newblock {\em arXiv e-prints}, page arXiv:2309.05050, September 2023.

\bibitem[NRJ24]{Nivesvivat:2023kfp}
Rongvoram Nivesvivat, Sylvain Ribault, and Jesper~Lykke Jacobsen.
\newblock {Critical loop models are exactly solvable}.
\newblock {\em SciPost Phys.}, 17:029, 2024.

\bibitem[Pol81]{polyakov1981quantum}
A.~M. Polyakov.
\newblock Quantum geometry of bosonic strings.
\newblock {\em Phys. Lett. B}, 103(3):207--210, 1981.

\bibitem[PT02]{PT02}
B.~Ponsot and J.~Teschner.
\newblock Boundary {L}iouville field theory: boundary three-point function.
\newblock {\em Nuclear Phys. B}, 622(1-2):309--327, 2002.

\bibitem[RZ22]{RZ22}
Guillaume Remy and Tunan Zhu.
\newblock Integrability of boundary {L}iouville conformal field theory.
\newblock {\em Comm. Math. Phys.}, 395(1):179--268, 2022.

\bibitem[Sab15]{saberi2015recent}
Abbas~Ali Saberi.
\newblock Recent advances in percolation theory and its applications.
\newblock {\em Phys. Rep.}, 578:1--32, 2015.

\bibitem[Sch00]{Sc00}
Oded Schramm.
\newblock Scaling limits of loop-erased random walks and uniform spanning trees.
\newblock {\em Israel J. Math.}, 118:221--288, 2000.

\bibitem[SD87]{SaleurDuplantier-1987}
H.~Saleur and B.~Duplantier.
\newblock Exact determination of the percolation hull exponent in two dimensions.
\newblock {\em Phys. Rev. Lett.}, 58(22):2325--2328, 1987.

\bibitem[She09]{SheffieldCLE}
Scott Sheffield.
\newblock Exploration trees and conformal loop ensembles.
\newblock {\em Duke Math. J.}, 147(1):79--129, 2009.

\bibitem[She16]{She16a}
Scott Sheffield.
\newblock Conformal weldings of random surfaces: {SLE} and the quantum gravity zipper.
\newblock {\em Ann. Probab.}, 44(5):3474--3545, 2016.

\bibitem[She23]{Sheffield-2022ICM}
Scott Sheffield.
\newblock What is a random surface?
\newblock In {\em I{CM}---{I}nternational {C}ongress of {M}athematicians. {V}ol. 2. {P}lenary lectures}, pages 1202--1258. EMS Press, Berlin, [2023] \copyright 2023.

\bibitem[Smi01]{Sm01}
Stanislav Smirnov.
\newblock Critical percolation in the plane: conformal invariance, {C}ardy's formula, scaling limits.
\newblock {\em C. R. Acad. Sci. Paris S\'{e}r. I Math.}, 333(3):239--244, 2001.

\bibitem[Smi10]{Smirnov-10}
Stanislav Smirnov.
\newblock Conformal invariance in random cluster models. {I}. {H}olomorphic fermions in the {I}sing model.
\newblock {\em Ann. of Math. (2)}, 172(2):1435--1467, 2010.

\bibitem[SSW09]{SSW09}
Oded Schramm, Scott Sheffield, and David~B. Wilson.
\newblock Conformal radii for conformal loop ensembles.
\newblock {\em Comm. Math. Phys.}, 288(1):43--53, 2009.

\bibitem[STZ{\etalchar{+}}22]{STZJ22}
Yu-Feng Song, Xiao-Jun Tan, Xin-Hang Zhang, Jesper~Lykke Jacobsen, Bernard Nienhuis, and Youjin Deng.
\newblock Nested closed paths in two-dimensional percolation.
\newblock {\em J. Phys. A}, 55(20):Paper No. 204002, 11, 2022.

\bibitem[SW01]{SW01}
Stanislav Smirnov and Wendelin Werner.
\newblock Critical exponents for two-dimensional percolation.
\newblock {\em Math. Res. Lett.}, 8(5-6):729--744, 2001.

\bibitem[SXZ24]{SXZ24-annulus}
Xin {Sun}, Shengjing {Xu}, and Zijie {Zhuang}.
\newblock {Annulus crossing formulae for critical planar percolation}.
\newblock {\em arXiv e-prints}, page arXiv:2410.04767, October 2024.

\bibitem[Tut63]{Tutte63}
W.~T. Tutte.
\newblock A census of planar maps.
\newblock {\em Canadian J. Math.}, 15:249--271, 1963.

\bibitem[{Wu}22]{Wu-annulus}
Baojun {Wu}.
\newblock {Conformal Bootstrap on the Annulus in Liouville CFT}.
\newblock {\em arXiv e-prints}, page arXiv:2203.11830, March 2022.

\bibitem[ZZ96]{ZZ96}
Alexander~B. Zamolodchikov and Alexei~B. Zamolodchikov.
\newblock {Structure constants and conformal bootstrap in Liouville field theory}.
\newblock {\em Nucl. Phys. B}, 477:577--605, 1996.

\end{thebibliography}

\end{document}